\documentclass[fleqn,usenatbib]{mnras}

\usepackage{mathrsfs}
\usepackage{graphicx}
\usepackage{subfigure}
\usepackage{epstopdf}
\usepackage{amsmath}
\usepackage{amssymb}
\usepackage{hyperref}
\usepackage{color}

\DeclareRobustCommand{\VAN}[3]{#2}
\let\VANthebibliography\thebibliography
\def\thebibliography{\DeclareRobustCommand{\VAN}[3]{##3}\VANthebibliography}

\title[Constraints on the abundance of SMPBH from lensing of CRS]{Constraints on the abundance of supermassive primordial black holes from lensing of compact radio sources}

\author[H. Zhou et al. 2022]{
Huan Zhou,$^{1}$ Yujie Lian,$^{2}$
Zhengxiang Li,$^{2}$\thanks{E-mail: zxli918@bnu.edu.cn}
Shuo Cao,$^{2}$\thanks{E-mail: caoshuo@bnu.edu.cn}
and Zhiqi Huang$^{1}$
\\
$^{1}$School of Physics and Astronomy, Sun Yat-sen University, Zhuhai, 519082, China\\
$^{2}$Department of Astronomy, Beijing Normal University, Beijing 100875, China\\
}

\date{Accepted 2022 March 29. Received 2022 March 1; in original form 2021 October 10}

\pubyear{2022}

\begin{document}
\label{firstpage}
\pagerange{\pageref{firstpage}--\pageref{lastpage}}
\maketitle

\begin{abstract}
The possibility that primordial black holes (PBHs) form a part of dark matter has been considered over a wide mass range from the Planck mass ($10^{-5}~\rm g$) to the level of the supermassive black hole in the center of the galaxy. Primordial origin might be one of the most important formation channel of supermassive black holes. We use the non-detection of lensing effect of very long baseline interferometer observations of compact radio sources with extremely high angular resolution as a promising probe to constrain the abundance of intergalactic PBHs in the mass range $\sim10^4$-$10^9~M_{\odot}$. For a sample of well-measured 543 flat-spectrum compact radio sources, no milli-lensed images are found with angular separations between $1.5$ milli-arcseconds and $50$ milli-arcseconds. From this null search result, we derive that the fraction of dark matter made up of supermassive PBHs in the mass range $\sim10^6$-$10^8~M_{\odot}$ is $\lesssim1.48\%$ at $95\%$ confidence level. This constraints would be significantly improved due to the rapid increase of the number of measured compact radio sources. For instance, on the basis of none confirmed milli-lensing candidate in the latest $\sim14000$ sources, we derive the abundance of supermassive PBHs and obtain that it is $\lesssim0.06\%$ at $95\%$ confidence level.
\end{abstract}

\begin{keywords}
Supermassive primordial black holes, Compact radio sources, Gravitational lensing
\end{keywords}

\section{Introduction}

The cosmological constant plus cold dark matter ($\Lambda$CDM) model has explained the evolution of the universe successfully. The scenario where CDM accounts for about a quarter of the total energy density is well consistent with current cosmological observations. However, we still know little about the constituent of dark matter (DM). Primordial black holes (PBHs)~\citep{Hawking1971,Carr1974,Carr1975}, which are predicted to form in the infant universe via different mechanisms, such as the enhanced curvature perturbations during inflation~\citep{Clesse2015,Pi2018,Ashoorioon2019,Fu2019,Cai2019,Motohashi2020}, bubble collisions~\citep{Hawking1982}, cosmic string~\citep{Hogan1984,Hawking1989}, and domain wall~\citep{Caldwell1996}, have been considered to be a promising candidate for the long elusive missing DM and therefore been a source of interest for nearly half a century. More interestingly, (stellar mass) PBHs have been attracting particular attention since the first detection of gravitational wave (GW) signal from the merger of black hole binary~\citep{Abbott2016}. This signal also can be interpreted as ripples of spacetime from the merger of PBH binary~\citep{Bird2016,Sasaki2016,Clesse2017a}.

So far, numerous methods have been proposed to constrain the abundance of PBHs, usually quoted as the fraction of PBHs in DM $f_{\rm PBH}=\Omega_{\rm PBH}/\Omega_{\rm DM}$ at present universe, in various possible mass windows including direct observational effects: gravitational lensing~\citep{Kassiola1991,Allsman2001,Wilkinson2001,Tisserand2007,Griest2013,Mediavilla2017,Zumalacarregui2018,Niikura2019,Niikura12019a,Liao2020,Zhou2021,Zhou2022}, dynamical effects on ultrafaint dwarf galaxies~\citep{Brandt2016,Koushiappas2017}, disruption of white dwarfs~\citep{Graham2015}, the effect of accretion via cosmic microwave background observations~\citep{Chen2016,Haimoud2017,Aloni2017,Poulin2017,Bernal2017}, nondetections of stochastic GW from binary black holes~\citep{Clesse2017b,Wang2018,Chen2020,Luca2020,Gert2020}, (extra)galactic $\gamma$-ray backgrounds~\citep{Carr2016,DeRocco2019,Laha2019,Laha2020a}, and indirect observational effects: null detection of scalar-induced GW~\citep{Chen2019b}, cosmic microwave background (CMB) spectral distortions from the primordial density perturbations~\citep{Carr1993,Carr1994}. In addition to these available probes, some other constraints from the near future observations, such as gravitational lensing of GW~\citep{Diego2020,Liao2020a,Urrutia2021}, gamma-ray bursts~\citep{Ji2018}, and 21 cm signals~\citep{Hektor2018,Clark2018,Halder2020}, have been proposed. See \citet{Sasaki2018,Green2020,Carr2021} for a recent review.

In addition to stellar or much smaller mass ranges, PBHs are also appealing for investigating the issue of supermassive black holes (SMBHs) observed at very high redshifts. For instance, observations of quasars at $z\geq6$ indicate that SMBH with masses greater than $\sim10^9~M_{\odot}$~\citep{Yang2020}, which are challenging to be formed via some astrophysical processes~\citep{Woods2019}. That is, the formation mechanism of SMBH is still a mystery in astrophysics. Actually, even assuming that the black hole continues the Eddington-limited accretion, it is nearly impossible for stellar-mass black holes ($\sim10-1000$ $M_{\odot}$) growing to reach the mass of $\sim10^9~M_{\odot}$ during the lifetime of the universe at $z\simeq6$. Moreover, it is also not yet clear if such an efficient accretion can persistently operate for cases from stellar-mass black holes to SMBHs within the Hubble time. Therefore, PBHs as progenitor of SMBHs are an alternative possibility and have been widely studied in the literature~\citep{Duechting2004,Kawasaki2012,Nakama2016,Hasegawa2018,Kawasaki2019,Kitajima2020}. It is known that if PBHs with masses $10^4-10^{13}~M_{\odot}$ are assumed to originate from the Gaussian primordial curvature perturbation, such primordial perturbations ineluctably result in the spectral distortion of the CMB which significantly exceeds the upper limit obtained by the COBE/FRIAS experiment~\citep{Kohri2014}. A scenario that predicts inevitable clustering of PBHs from highly non-Gaussian perturbations has been proposed to create PBHs~\citep{Kohri2014,Nakama2016,Huang2019,Shinohara2020}. Therefore, direct probe for PBHs in the mass range $\sim10^4-10^9~M_{\odot}$ using the lensing effect of very long baseline interferometer (VLBI) observations of compact radio source (CRS) with extremely high angular resolution would be a complementary limit to the constraint of distortion of the CMB. In addition, ~\citet{Banik2019} have studied the effect of wandering black holes on VLBI images of lensed arcs.

In this paper, we first apply the method of optical depth to constrain the abundance of PBHs with well-measured 543 CRSs, observed by a well-known VLBI survey undertaken by~\citet{Preston85} (hereafter P85). Benefit from the high-quality maps obtained by VLBI, one is able to measure the milli-arcsecond ultra-compact structure in radio sources~\citep{Kellermann93} and search possible examples of multiple images produced by milli-lensing. Based on the the null search result with updated redshift measurements of P85~\citep{Jackson06}, we obtain the most stringent limit on the abundance of spuermassive primordial black holes (SMPBHs) in the mass range $\sim10^6-10^8~M_{\odot}$. In addition, we yield stronger constraints based on a larger catalog from ~\citet{Casadio2021}.

This paper is organized as follows: we introduce the CRS data and the theory of optical depth of CRS lensing in Section 2. In Section 3, we apply this method to the CRS observations and yield results. Conclusions and discussions are presented in Section 4. Throughout, we use the concordance $\Lambda$CDM cosmology with the best-fit parameters from the recent Planck observations~\citep{Planck2018}.

\section{Methodology}

In this section, we introduce the current data of CRS observations and specify the optical depth theory of CRS lensing based on the method from~\citet{Press1973, Kassiola1991, Wilkinson2001}.

\subsection{Compact Radio Source Observations}

\begin{figure}
    \centering
     \includegraphics[width=0.5\textwidth, height=0.36\textwidth]{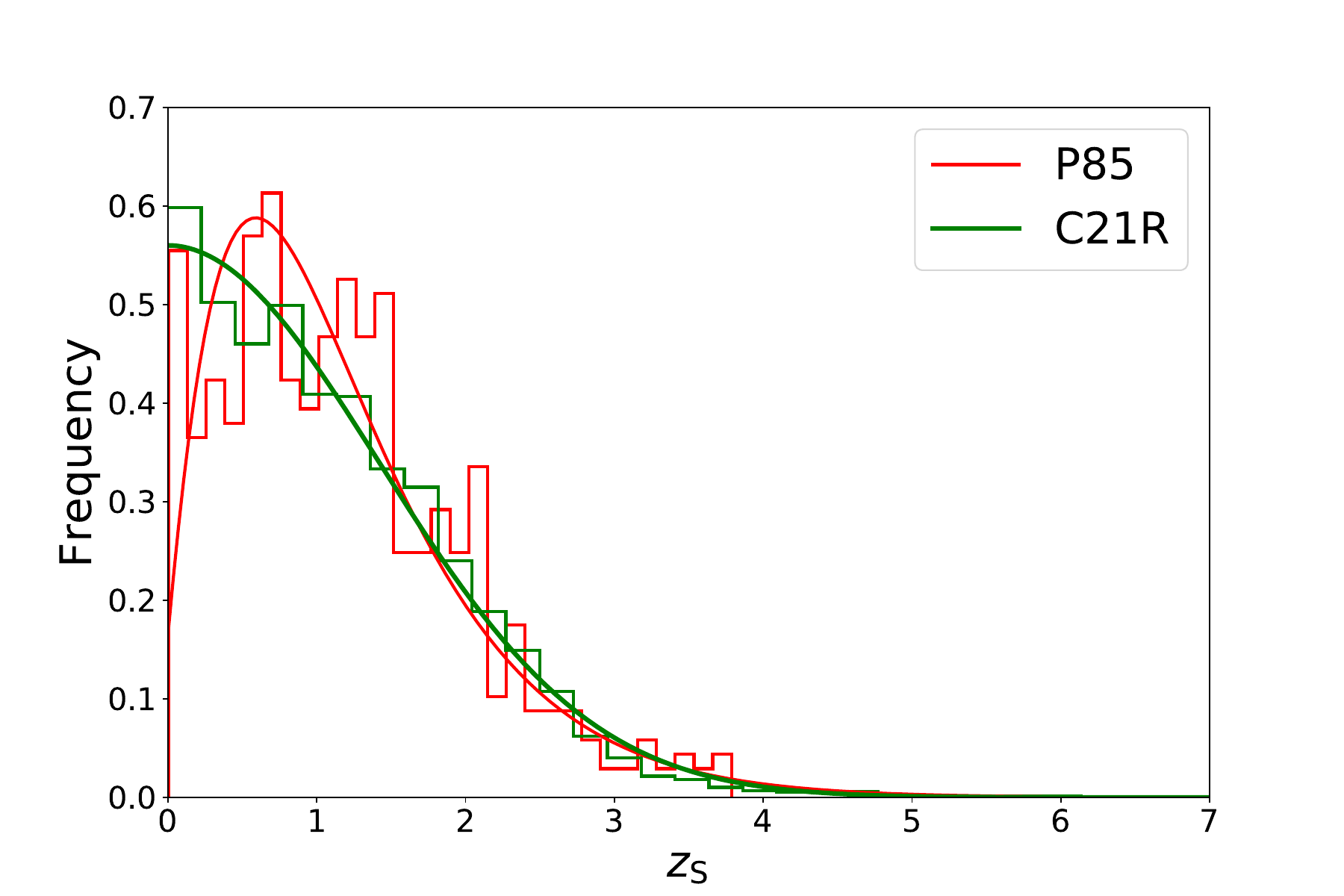}
     \caption{Red line represent the redshift distribution of well-measured 543 CRSs. Greed line represent the redshift distrbution from \citet{Casadio2021} CRSs with redshift information. }\label{fig1}
\end{figure}

The parent data used in this paper was derive from a catalog of ultra-compact radio sources, based on the observations of a 2.29 GHz VLBI all-sky survey~\citep{Preston85}. By employing a intercontinental VLBI array with an effective baseline of $\sim 8\times 10^7$ wavelengths, 917 sources have been systematically observed with compact structure out of 1398 known radio sources. Note that these detected extragalactic objects coincide with different optical counterparts such as BL Lac objects, quasars, and radio galaxies~\citep{Gurvits94,Gurvits99,Cao15,Cao17a,Cao17b,Cao18}, with a correlated flux limit of approximately 0.1 Jy. In this analysis, we focus on a revised sample including a significant fraction of P85 catalog, with updated redshift measurements and radio information for 613 objects covering the redshift range of $0.0035\leq z\leq 3.787$. The full description of the observations and the corresponding details (i.e., source name, angular size of the compact structure, total radio flux density, spectral index, and optical counterpart) can be found in~\citep{Preston85,Jackson06}~\footnote{A full list is available via http://nrl.northumbria.ac.uk/13109/.}. Such data has been extensively investigated in a number of cosmological studies, focusing on its possibility of establishing a sample of standard cosmological rulers at higher redshifts and exploring the evolution of early universe~\citep{Jackson04,Zheng17,Xu18,Qi19,Cao20,Zheng20,Liu21,Qi21}. On the other hand, it is well-known that radio-loud active galactic nuclei (AGN) typically contains a flat-spectrum core and a steep-spectrum jet. In the framework of core-jet model proposed by ~\citet{Blandford79}, flat-spectrum radio sources could be explained as the apparent origin of AGN jets, which always appear as the brightest unresolved compact cores in the VLBI images. Therefore, following the analysis of~\citet{Gurvits99}, we also apply a selection criterion to define the so-called ``flat-spectrum core", with spectral index $\alpha\geq-0.5$. In this context, the spectra index is defined as $S\propto\nu^\alpha$, where $S$ is the flux density and $\nu$ is the frequency. The redshift distribution for the final CRS sample ($N=543$), which contains all of the strongest flat-spectrum compact radio sources in P85 VLBI survey, is shown in Fig.~\ref{fig1}.

In order to carry out systematic search for strong gravitational lenses on mas-scales, with possible multiple images produced by gravitational lensing, we turn to the Astrogeo VLBI FITS image database
\footnote{http://astrogeo.org/vlbi\_images/.}, the NASA/IPAC Extragalactic
Database (NED) \footnote{http://ned.ipac.caltech.edu/.}, and the
VLBI-derived catalogue OCARS \footnote{http://www.gaoran.ru/english/as/ac\_vlbi/ocars.txt.}
\citep{Malkin2018} to obtain the radio images and additional information of these 543 flat-spectrum CRSs. Specially, the B1950 and J2000 name for the P85 radio sources are determined at the Astrogeo VLBI FITS image database at the time this search, with the source name, positions and redshift given by \citet{Preston85}. It is worth noting that the final CRS sample ($N=543$) were observed at multiple radio frequencies, showing an unresolved flat-spectrum core with high resolution in at least one of the available observing band. The presence of two or more flat-spectrum core images provides a simple and interesting probe to detect candidates of lensed CRSs, with massive PBHs acting as point-like lenses. We try to find suitable lens candidates from the 543 flat-spectrum CRSs, based on the three major selection criteria as mentioned in~\citet{Wilkinson2001}: I) Although the effect of lensing magnifies the brightness of the images, the total surface brightness of the source is conserved based on conservation of energy. For the surface-brightness distribution of CRS, the primary compact component should be much larger than the secondary and other counterparts; II) For the flux density of the CRS, the ratio of the primary and secondary compact components should be $\leq 40:1$; III) For the positions of the CRS cores, the separation of the primary and secondary compact components should be $\delta \leq \Delta \theta \leq \Delta$. Here $\delta$ and $\Delta$ denote the minimum and maximum image separations, i.e., the angular resolution and the limited field of view (FoV) of the observation with which each radio source was observed. In this analysis we make a conservative estimation and take the typical value of $\delta=1.5$ mas (the achievable resolution with the visibility distribution of the VLBI core) and $\Delta=50$ mas (the field of view of $\sim 50\times50$ mas) for the P85 radio sources \citep{Wilkinson2001}.

At the first stage, we visually inspect all of the radio sources in all available frequencies searching for images with multiple compact components on mas-scales. After a careful check of the final CRS sample, 43 sources were selected, showing multiple compact components in at least one of the available bands. Now we report a few extra notes and considerations on all of the possible lensed sources. I) 31 candidates are confirmed to be core-jet sources, where the radio image exhibits a bright core and a faint jet with lower brightness hot-spot. This directly allows us to discard them as a lenses. II) 3 candidates turn out to be Compact Symmetric Objects (CSOs), which often show a single central core straddled by a pair of outer lobes of significantly lower surface brightness \citep{Fanti1995,Readhead1996}. The corresponding orders of these three sources in P85 catalog are 982, 1062, and 1076, while the J2000 names of these sources are J1558-1409, J1723-6500, and J1734+0926 \citep{Sokolovsky2011,Angioni2019,An2012}. It is interesting to note that J1558-1409 has been identified as a CSO candidate but not a confirmed one \citep{Sokolovsky2011}. However, compared with the 2 GHz radio images, its 8 GHz radio image reveals a new component between the two main ones, suggesting that the structure consists of a bright core and knots in a faint underlying lobe. III) 4 sources J0956+2515, J1357+4353, J2247-1237, and J2354-0019, whose orders are 567, 865, 1308, and 1384 in P85 catalog are rejected on account of at least one of the three basic selection criteria. For instance, J1357+4353 has also been identified as a CSO candidate by the recent work of \citet{Sokolovsky2011}, in which the flux density ratios between the two compact component ($1.512 \pm 0.111$ at 2 GHz and $2.208 \pm 0.235$ at 8 GHz) were reported. Considering the criterion that the lensed images should have identical radio spectra, or equivalently consistent flux ratio at different frequencies, such source is excluded as a lens because of its inconsistent multi-frequency flux ratios. IV) For 4 candidates: J0008-2339, J1101+7225, J1809+2758, and J1939-1002 (whose orders are 9, 647, 1105, and 1148 in P85 catalog), the weaker secondary component is not detected at a different observing frequency, or its surface brightness is much lower than the compact primary counterpart. Although the nature of the secondary component is still a controversy, it is less possible to be a the other lensed image of compact component on mas-scale. V) For the source J1107-6820 with the order of 661 in P85 catalog, \citet{Ojha2005} has reported the VLBI 8.4 GHz radio image and flux density ratio between the two compact components of J1107-6820. However, with the absence of radio images and flux density measurements at different frequencies, we fail to confirm its possibility to pass the follow-up tests and act as a milli-lensed candidate. Therefore, no strong evidence of milli-lensed candidate is found for the 543 flat-spectrum CRSs. Such conclusion is well consistent with the recent results of \citet{Casadio2021}, focusing on the search of probable milli-lensed candidates based on a sample of more than 13000 individual sources from Astrogeo. It is worth noting that the two candidates (J0525+1743, J2312+0919), with a higher probability of being associated with gravitational lenses are not included in our P85 CRS sample. In addition, we use the larger VLBI catalog with the FoV of $150$ mas from \citet{Casadio2021} as a comparison. In the latest and largest sample, there are about $8000$ sources with redshift information which is from the footnote 4, and the distribution of their redshifts is plotted in Figure~\ref{fig1}. In following analysis, we named the full sample of $\geq13000$ sources from \citet{Casadio2021} as `C21' and denoted the subsample of sources with redshift information as `C21R'. Moreover, we also assume the redshift distribution of C21 is the same as the one of C21.

\subsection{Lensing of Compact Radio Sources}
\begin{figure}
    \centering
     \includegraphics[width=0.5\textwidth, height=0.36\textwidth]{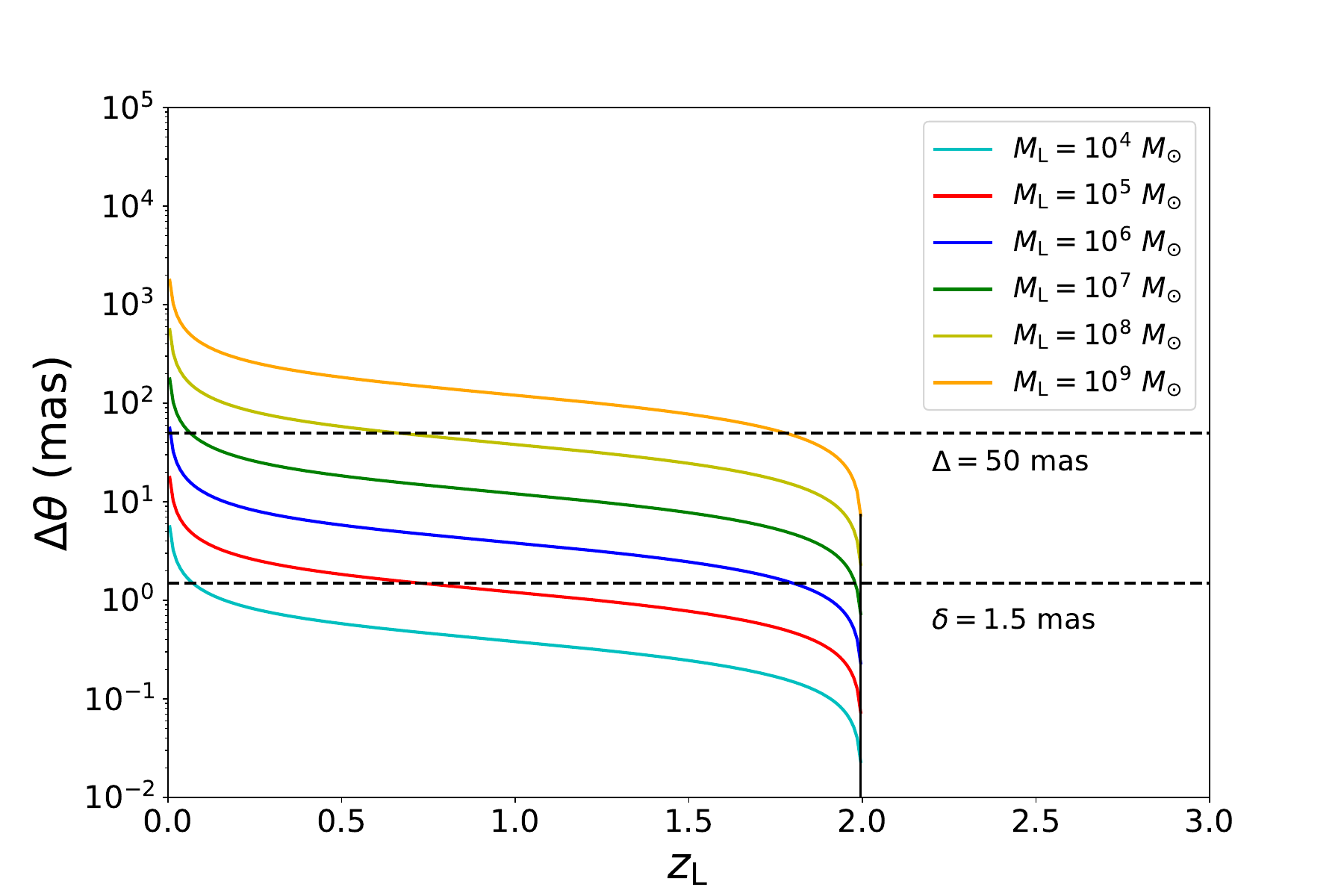}
     \caption{The angular separation of two images as a function with respect to the redshift of the lens, for a source at redshift $z_{\rm S}=2$. The mass of the lens ranges from $10^4$ to $10^9~M_{\odot}$. The magnifications ratio of the images is $R_{\rm f}=40$.}\label{fig2}
\end{figure}

For a point mass $M_{\rm PBH}$ lens, Einstein radius is
\begin{equation}\label{eq2}
\theta_{\rm E}=2\sqrt{\frac{GM_{\rm PBH}}{c^2D}}\approx (1~\rm mas)\bigg(\frac{M_{\rm PBH}}{10^5M_{\odot}}\bigg)^{1/2}\bigg(\frac{D}{\rm Gpc}\bigg)^{-1/2},
\end{equation}
where $G$ and $c$ denote the gravitational constant and the speed of light, respectively. In addition, $D=D_{\rm L}D_{\rm S}/D_{\rm LS}$ is effective lensing distance, where $D_{\rm S}$, $D_{\rm L}$, and $D_{\rm LS}$ represent the angular diameter distance to the source, to the lens, and between the source and the lens, respectively. The angular resolution for some CRSs with very long baseline array (VLBA) could reach a high level, e.g. $\sim\rm mas$, it is possible to distinguish multiple images of a CRS lensed by intervening objects with mass greater than $\sim10^5M_{\odot}$. Because the Schwarzschild radius of PBH is much smaller than the Einstein radius multiplied by the angular diameter distance of the lens, i.e. $R_{\rm PBH}=2GM_{\rm PBH}/c^2\ll\theta_{\rm E}D_{\rm L}$, the lens equation of a point mass $M_{\rm PBH}$ is
\begin{equation}\label{eq3}
\theta^2-\beta\theta-\theta_{\rm E}^2=0,
\end{equation}
where $\beta$ stands for the source position. The above lens equation of a point mass lens means there would be two images at positions
\begin{equation}\label{eq4}
\theta_{\pm}=\frac{1}{2}\bigg(\beta\pm\sqrt{\beta^2+4\theta_{\rm E}^2}\bigg).
\end{equation}
It is obvious that one image locates outside the Einstein ring with $\theta_{+}>\theta_{\rm E}$ and the other one is within the ring $\theta_{-}<\theta_{\rm E}$. Their magnifications $\mu_{\pm}$ satisfy
\begin{equation}\label{eq5}
\mu_{\pm}=\bigg(1-\frac{\theta_{\rm E}^4}{\theta_{\pm}^4}\bigg)^{-1}.
\end{equation}
In addition, the magnification ratio between two images can be directly obtained as well,
\begin{equation}\label{eq6}
R_{\rm f}\equiv\bigg|\frac{\mu_{+}}{\mu_{-}}\bigg|=\bigg|\frac{\theta_{+}}{\theta_{\rm E}}\bigg|^4,
\end{equation}
where the relation between two images $\theta_{+}\theta_{-}=-\theta_{\rm E}^2$ is used. The normalized impact parameter $y\equiv\frac{\beta}{\theta_{\rm E}}$ for a reference value of magnification ratio $R_{\rm f}$ is
\begin{equation}\label{eq7}
y(R_{\rm f})=R_{\rm f}^{1/4}-R_{\rm f}^{-1/4}.
\end{equation}
In order to make both lensed images, especially the fainter one, detectable with high enough signal-noise ratio, $R_{\rm f}$ should be smaller than a threshold $R_{\rm f,max}$. Following ~\cite{Wilkinson2001}, we set the maximum value of magnification ratio $R_{\rm f,max}=40$. This value is reasonable for persistent radio sources, e.g. CRSs.

The lensing cross section due to a PBH lens is given by
\begin{equation}\label{eq8}
\begin{split}
\sigma(M_{\rm PBH}, z_{\rm L}, z_{\rm S})=\frac{4\pi GM_{\rm PBH}D_{\rm L}D_{\rm LS}}{c^2D_{\rm S}}[y^2_{\rm max}(\Delta,M_{\rm PBH},z_{\rm L},z_{\rm S})\\
-y^2_{\rm min}(\delta,M_{\rm PBH},z_{\rm L},z_{\rm S})],
\end{split}
\end{equation}
where the maximum impact parameter $y_{\rm max}(\Delta,M_{\rm PBH},z_{\rm L},z_{\rm S})$ and minimum impact parameter $y_{\rm min}(\delta,M_{\rm PBH},z_{\rm L},z_{\rm S})$ are determined by the angular resolution $\delta$ and the FoV of the observation $\Delta$, respectively. Meanwhile, the impact parameter must be smaller than $y_{\rm max}(R_{\rm f,max})$. Now, we explain the maximum and minimum impact parameter in the cross section. The angular separation of two images lensed by a PBH is
\begin{equation}\label{eq9}
\Delta\theta=\theta_{\rm E}(R_{\rm f}^{1/4}+R_{\rm f}^{-1/4}).
\end{equation}
Illustratively, as shown in Fig.~\ref{fig2}, for a CRS at redshift $z_{\rm S}=2$, imaged with a fixed $R_{\rm f}=40$, $\delta=1.5~\rm mas$, and $\Delta=50~\rm mas$, we have plotted the angular separation of the images against the redshift of the lens with different masses. It is obvious that the angular separation $\Delta\theta$ decreases as lens redshfit $z_{\rm L}$ increases. 
For a source at $z_{\rm S}=2$ and small mass PBH lenses ($\lesssim10^6~M_{\odot}$), the upper limit of the lens redshift producing two detectable images is truncated by the the angular resolution $\delta$ (the lower dotted line in Fig.~\ref{fig2}). While for large mass PBH lenses ($\gtrsim10^7~M_{\odot}$), the lower limit of the lens redshift producing two detectable images is determined by the the FoV $\Delta$ (the upper dotted line in Fig.~\ref{fig2}). It is obvious that the upper bound of lens redshift is very close to the one of the source and lower bound is very close to zero within the mass range $10^6\sim10^7~M_{\odot}$. 

If the angular resolution is $\delta$, we only are able to detect the presence of lens mass $M_{\rm PBH}$ assuming a CRS is at redshift $z_{\rm S}$ when the value of angular separation of two images satisfy 
\begin{equation}\label{eq10}
\Delta\theta\geq\delta. 
\end{equation}
This requirement results in a minimum impact parameter $y_{\rm min}(\delta,M_{\rm PBH},z_{\rm L},z_{\rm S})$. Analogously, we can only detect secondary images if they lie within the FoV of the observation, $\Delta$. That is, the value of angular separation of two images must satisfy
\begin{equation}\label{eq11}
\Delta\theta\leq\Delta.
\end{equation}
This condition yields a maximum impact parameter $y_{\rm max}(\Delta,M_{\rm PBH},z_{\rm L},z_{\rm S})$. For a given lens mass $M_{\rm PBH}$ and a CRS at redshift $z_{\rm S}$, equation~(\ref{eq11}) will always be violated when the redshift of lens is smaller than $z_{\Delta}$, a solution of  $\Delta\theta=2\theta_{\rm E}=\Delta$. This violation heralds that the minimum angular separation of two images is larger than the FoV of the observation. Therefore, there is no contribution of cross section at redshifts smaller than $z_{\Delta}$. Moreover, when the redshift of lens is between $z_{\Delta}$ and $z_{\delta}$ that comes from the $\Delta\theta=2\theta_{\rm E}=\delta$, the minimum angular separation of two images come in the range of FoV but is still larger than the angular resolution. In this case, the minimum impact parameter $y_{\rm min}(\delta,M_{\rm PBH},z_{\rm L},z_{\rm S})$ must be zero. However, the maximum impact parameter $y_{\rm max}(\Delta,M_{\rm PBH},z_{\rm L},z_{\rm S})$ can be derived from equation~(\ref{eq11}) as follows
\begin{equation}\label{eq12}
\begin{split}
y_{\rm max}(\Delta,M_{\rm PBH},z_{\rm L},z_{\rm S})=
\frac{\Delta/\theta_{\rm E}+\sqrt{(\Delta/\theta_{\rm E})^2-4}}{2}-\\
\frac{2}{\Delta/\theta_{\rm E}+\sqrt{(\Delta/\theta_{\rm E})^2-4}}.
\end{split}
\end{equation}
When the redshift of lens is lager than $z_{\delta}$, the minimum impact parameter $y_{\rm min}(\delta,M_{\rm PBH},z_{\rm L},z_{\rm S})$ can be derived equation~(\ref{eq9}) as 
\begin{equation}\label{eq13}
\begin{split}
y_{\rm min}(\delta,M_{\rm PBH},z_{\rm L},z_{\rm S})=
\frac{\delta/\theta_{\rm E}+\sqrt{(\delta/\theta_{\rm E})^2-4}}{2}-\\
\frac{2}{\delta/\theta_{\rm E}+\sqrt{(\delta/\theta_{\rm E})^2-4}}.
\end{split}
\end{equation}
It should be pointed out that the method in ~\cite{Kassiola1991} for calculating the truncated redshifts leads to underestimate of cross section contribution from the low redshift and the redshift close to the source. Therefore, for a single source, the lensing optical depth due to a single PBH lens should be
\begin{equation}\label{eq14}
\begin{split}
\tau(M_{\rm PBH},f_{\rm PBH},z_{\rm S})=\int_{0}^{z_{\rm S}}d\chi(z_{\rm L})
(1+z_{\rm L})^2\sigma(M_{\rm PBH},z_{\rm L},z_{\rm S})\\
\times n_{\rm L}(f_{\rm PBH})=\frac{3}{2}f_{\rm PBH}\Omega_{\rm DM}\int_{0}^{z_{\rm S}}dz_{\rm L}\frac{H_0^2}{cH(z_{\rm L})}\frac{D_{\rm L}D_{\rm LS}}{D_{\rm S}}(1+z_{\rm L})^2\\
\times[y^2_{\rm max}(\Delta,M_{\rm PBH},z_{\rm L},z_{\rm S})-y^2_{\rm min}(\delta,M_{\rm PBH},z_{\rm L},z_{\rm S})],
\end{split}
\end{equation}
where $n_{\rm L}$ is the comoving number density of the lens, $H_0$ is the Hubble constant, $H(z_{\rm L})$ is the Hubble parameter at $z_{\rm L}$, and $\Omega_{\rm DM}$ is the present fractional density of DM.

Now, for a given distribution function $N(z_{\rm S})$ of CRSs, their integrated optical depth $\bar{\tau}(M_{\rm PBH},f_{\rm PBH})$ is 
\begin{equation}\label{eq15}
\bar{\tau}(M_{\rm PBH},f_{\rm PBH})=\int dz_{\rm S}\tau(M_{\rm PBH},f_{\rm PBH},z_{\rm S})N(z_{\rm S}).
\end{equation}
If one observes a large number of CRSs, $N_{\rm CRS}$, then the number of detectable lensed CRSs is expected to be
\begin{equation}\label{eq16}
N_{\rm lensed~CRS}=(1-e^{-\bar{\tau}(M_{\rm PBH},f_{\rm PBH})})N_{\rm CRS}.
\end{equation}
If none of the CRS is found to be lensed, then the constraint on the upper limit of the fraction of DM in the form of PBHs can be estimated from equation~(\ref{eq16}).

\section{Results}
\begin{figure}
    \centering
     \includegraphics[width=0.5\textwidth, height=0.36\textwidth]{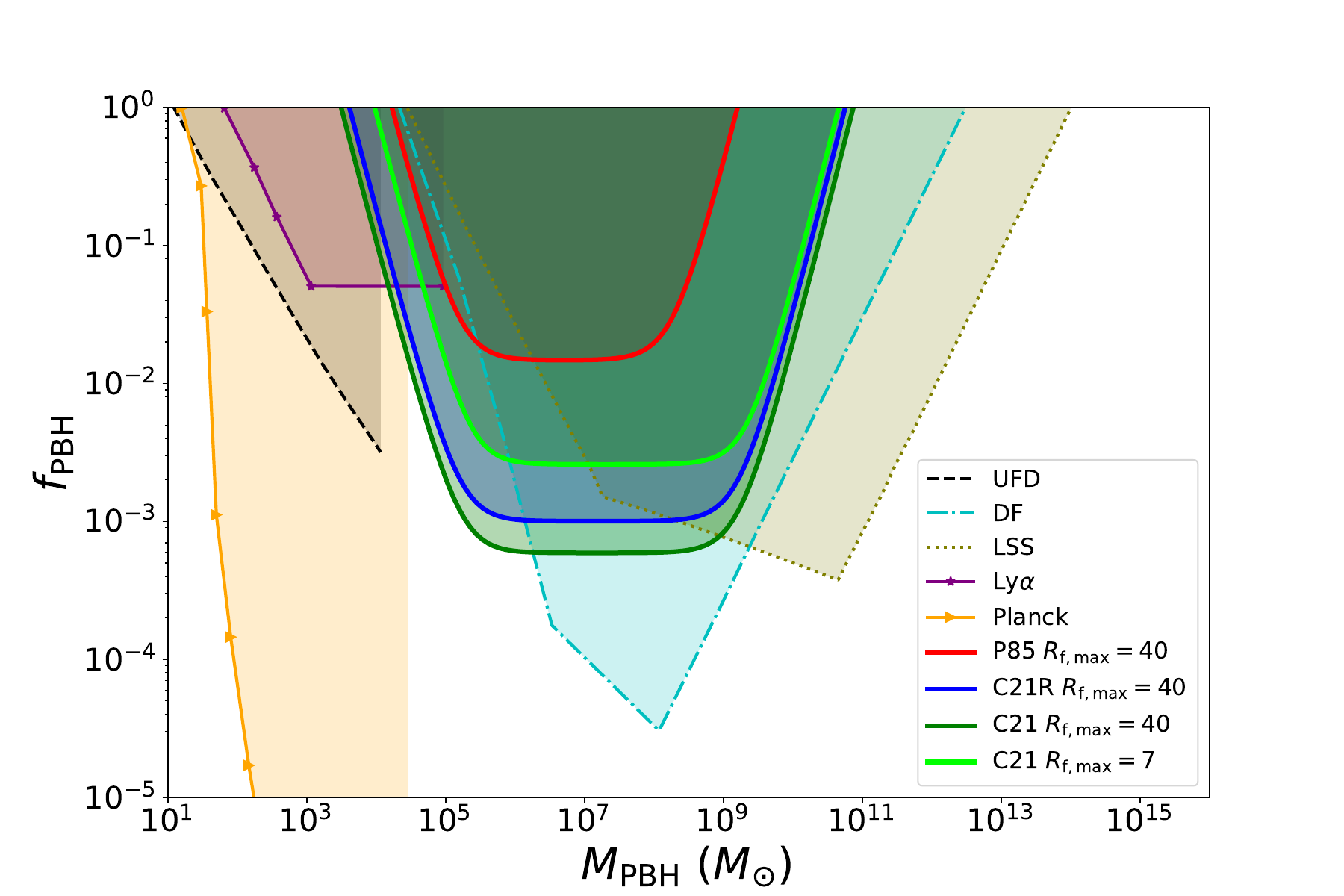}
     \caption{Red solid line represents the constraint on $f_{\rm PBH}$ at the $95\%$ confidence level from the null search result of lensing candidate in the well-measured P85 sources with image separations from 1.5 to 50 mas. Blue solid line and green solid line represent the constraints on $f_{\rm PBH}$ at the $95\%$ confidence level from the null search result of lensing candidate in the C21R and C21 catalogs with image separations ranging from 1.5 to 150 mas, respectively. Lime solid line represents the constraint on $f_{\rm PBH}$ with $R_{\rm f,max}=7$ in the C21 catalog. Other constraints include the ultra-faint dwarfs (UFD)~\citep{Brandt2016}, infalling of halo objects due to dynamical friction (DF)~\citep{Carr1999}, various cosmic large-scale structure (LSS)~\citep{Carr2018}, Lyman-$\alpha$ forest observations (Ly$\alpha$)~\citep{Murgia2019}, CMB anisotropy measured by Planck (Planck)~\citep{Serpico2020}.}\label{fig3}
\end{figure}

We first use the optical depth method to constrain the abundance of SMPBHs. Each source is observed with an approximately fixed angular resolution corresponding to the minimum image separation, and limited FoV. The angular resolution and the limited FoV act to truncate the mass range of the lenses that produce detectable multiple images. The minimum value of upper limit of the abundance of SMPBHs is insensitive to the angular resolution and FoV. This inference is similar as the one concluded in~\citet{Kassiola1991}. If there is no truncation on mass, the comoving number density $n$ of PBH with a particular mass is proportional to $1/M_{\rm PBH}$ for a given value of $f_{\rm PBH}$. In addition, the gravitational lensing cross section $\sigma$ is $\propto M_{\rm PBH}$, and hence the path length to lensing $1/(n\sigma)$ is independent on the lens mass. Thus, the optical depth across the measurable mass range is roughly a constant, yielding a universal upper limit of $f_{\rm PBH}$ which is proportional to $y^2(R_{\rm f,max})$ (equation~(\ref{eq7})). Compared with the common definition of strong lensing referring to images with a magnification ratio $R_{\rm f}=7$~\citep{Turner1984}, the $R_{\rm f,max}=40$ magnification ratio corresponds to a configuration with a larger impact parameter and hence the lensing optical depth (cross section) is almost 4 times larger than that normally assumed for lensing calculations. Therefore, as shown in Figure~\ref{fig3}, the $R_{\rm f,max}=40$ would result in about 4 times stronger upper limits of $f_{\rm PBH}$ than the constraint with $R_{\rm f,max}=7$. For 543 well-measured CRSs from P85, our search is sensitive to the mass range from $\sim10^6~M_{\odot}$ to $\sim10^8~M_{\odot}$ with $\delta=1.5~\rm mas$ and $\Delta=50~\rm mas$. In addition, within this lens mass range, the null search result of lensed CRSs leads to a constraint on the upper limit of $f_{\rm PBH}$. As shown in Fig.~\ref{fig3}, at the $95\%$ confidence level, the fraction of DM in the form of SMPBHs with the mass ranging from $\sim10^6~M_{\odot}$ to $\sim10^8~M_{\odot}$ is $\lesssim1.48\%$. 

In addition to 543 well-measured sources from P85, we also consider the latest full sample from~\citet{Casadio2021} with the minimum image separation, $\delta=1.5~\rm mas$, and limited FoV, $\Delta=150~\rm mas$. Assuming none confirmed lensed CRS in C21R and C21, we derive that the upper limit of $f_{\rm PBH}$ in the mass range $\sim10^6$-$10^9~M_{\odot}$ is $\lesssim0.10\%$ and $\lesssim0.06\%$ at $95\%$ confidence level, respectively. If the two candidates with a higher probability are verified as gravitational lensing systems with $10^7~M_{\odot}$ lens~\citet{Casadio2021}, the $f_{\rm PBH}$ will be inferred as $0.04\%$. However, these candidates require further validation by future follow-up observations. For all three samples (P85, C21R, C21), we find that the constraints scale show consistent power law, $f\propto (m\cdot M_{\odot})^{-1.9}$ and $f\propto (m\cdot M_{\odot})^{1.9}$, in low and high mass ends, respectively. Moreover, although this asymptotic scale is neither sensitive to angular resolution nor to FoV, the positions where the asymptotic scale occurs are dependent on them. Therefore, future observations with higher angular resolution and larger FoV will be able to yield constraints on a wider mass range of PBHs.

\begin{figure}
    \centering
     \includegraphics[width=0.5\textwidth, height=0.36\textwidth]{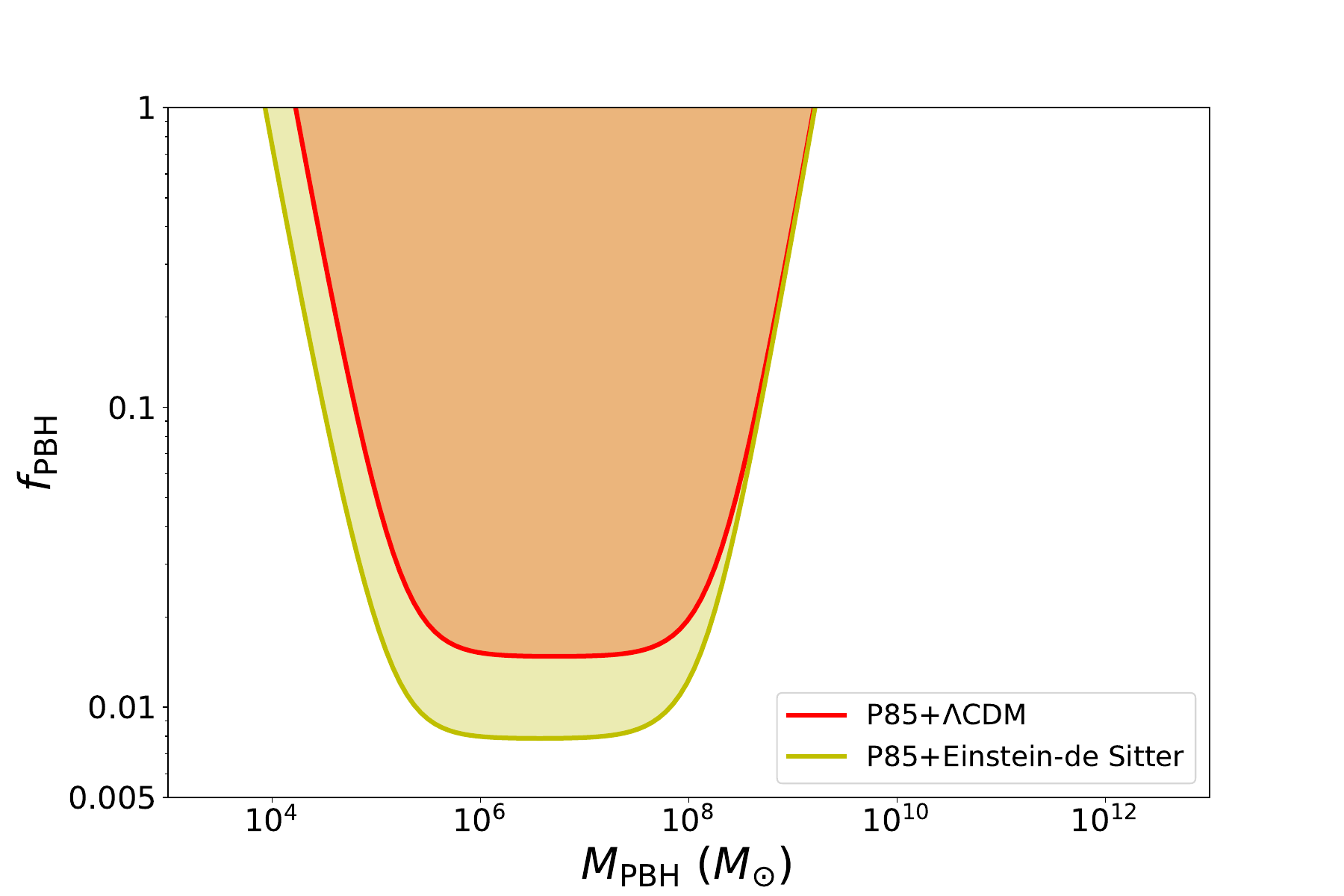}
     \caption{For the P85 catalog, red line and yellow line represent the constraint on $f_{\rm PBH}$ with $R_{\rm f,max}=40$ at the $95\%$ confidence level based on the $\Lambda$CDM model and  the Einstein-de Sitter model, respectively.}\label{fig4}
\end{figure}

Compared with the upper limit on the abundance of compact objects presented in~\citet{Wilkinson2001}, our constraints have been significantly improved. The improvement firstly arises from the optical depth method which yields more reasonable estimation for cross section contribution. This method allows a wider mass range of SMPBHs to be limited compared with the method in~\citet{Kassiola1991}, but does not affect the minimum of $f_{\rm PBH}$ since the upper limit of the abundance of SMPBHs is insensitive to $\delta$ and $\Delta$ at the middle mass range. Secondly, the improvement partly comes from the use the standard $Planck$ best-fit $\Lambda$CDM model. As shown in Figure~\ref{fig4}, the Einstein-de Sitter universe ($\Omega_{\rm M}=1, \Omega_{\Lambda}=0$) used in~\citet{Wilkinson2001} would result in about 2 times stronger upper limits of $f_{\rm PBH}$. The final improvement is from the increase of the number of sources and it is the main factor ($\sim500$, $\sim8000$, $\sim14000$, cf. 300). Assuming all samples follow the same redshift distribution, the increase of sample size of P85, C21R, and C21 would lead to 2 times, 27 times, and 46 times stronger upper limits of $f_{\rm PBH}$ than the real result presented in ~\citet{Wilkinson2001}, respectively. These improved constraints would be helpful and complementary for probing the possibility of SMPBHs in this mass window making up DM.

\section{Conclusions and discussions}
In this paper, we have derived constraints on the presence of intergalactic PBHs in the mass range $\sim10^4-10^9~M_{\odot}$ with the lensing effect of CRS which are observed by VLBI with extremely high angular resolution. By searching examples of multiple images produced by milli-lensing in a sample of well-measured 543 CRSs, no evidence of lensed candidates was found with angular separations in the range 1.5-50 milli-arcseconds. Based on the optical depth method, we derive the $95\%$ confidence-level constraint on the fractional abundance of SMPBHs, $f_{\rm PBH}<1.48\%$, in the mass range $\sim10^6-10^8~M_{\odot}$. This constraint might be conservative, because gravitational lensing effect increases the observed flux density of a background source. As a result, lensed sources are drawn from a fainter source population than the unlensed sources. In this case, a flux-limited survey will probably contain more lensed candidates than expected. Therefore, the null search result formally leads to a more stringent bound on $f_{\rm PBH}$. This ``magnification bias” associated with sources used in our analysis is of order unity.

We expect the constraint on $f_{\rm PBH}$ to be significantly improved based on a larger catalog of CRS. Assuming detection of $\sim14000$ unlensed CRSs with same angular resolution and larger FoV~\citep{Casadio2021}, we find that uniformly distributed SMPBHs in the mass range from $\sim10^6~M_{\odot}$ to $\sim10^9~M_{\odot}$ do not make up more than $\sim0.06\%$ of DM at the $95\%$ confidence level.

The above direct constraints on SMPBH can provide clues about the origin of SMBHs. It is well known that the formation of black holes with masses greater than $\sim10^9~M_{\odot}$ observed at $z\geq6$ is still mysterious since the continuous Eddington-limited accretion is far insufficient for stellar-mass black holes growing to SMBHs during the life-time of the universe. PBHs have been widely proposed as seeds of SMBHs. Therefore, direct constraints from the well-measured 543 CRSs would be helpful for exploring this issue. Moreover, SMPBHs can form binaries, coalesce, and produce GWs in nano-Hertz frequency band, which can be detected by using stable millisecond pulsars~\citep{Jaffe2003,Sesana2008,Sesana2009}. The null-detection of GWs by pulsar timing array (PTA) can successfully constrain continuous GWs from individual supermassive binaries black holes~\citep{Zhu2014,Babak2016,Aggarwal2019}. It suggests that PTA experiment also is an important tool to constrain SMBHs in the near future. In combination of milli-lensing of CRSs and measurment of continuous GWs from individual supermassive binaries black holes, it is foreseen that upcoming complementary multi-messenger observations will yield considerable constraints on the possibilities of SMPBHs.

\section*{Acknowledgements}
This work was supported by the National Natural Science Foundation of China under Grants Nos. 11920101003, 11722324, 11603003, 11633001, 12073088, and U1831122; National Key R\&D Program of China No. 2017YFA0402600; Guangdong Major Project of Basic and Applied Basic Research (Grant No. 2019B030302001); the Strategic Priority Research Program of the Chinese Academy of Sciences, Grant No. XDB23040100, and the Interdiscipline Research Funds of Beijing Normal University.

\section*{Data Availability}
The data underlying this article will be shared on reasonable request to the corresponding author.

\label{lastpage}

\begin{thebibliography}{}
\bibitem[Abbott et al.(2016)]{Abbott2016} Abbott, B. P., Abbott, R., Abbott, T. D., et al., 2016, PRL, 116, 061102

\bibitem[Aggarwal et al.(2019)]{Aggarwal2019}Aggarwal, K., Arzoumanian, Z., Baker, P. T., et al., 2019, ApJ, 880, 116

\bibitem[Planck Collaboration, (2020)]{Planck2018}Aghanim, N., Akrami, Y., Ashdown, M., et al., 2020, A\&A 641, A6

\bibitem[Alcock et al.(2001)]{Allsman2001}Alcock, C., Allsman, R. A., Alves, D. R., et al., 2001, ApJ, 550, L169

\bibitem[Ali-Ha\"{\i}moud, \& Kamionkowski(2017)]{Haimoud2017}  Ali-Ha\"{\i}moud Y.,  Kamionkowski M., 2017, PhRvD, 95, 043534

\bibitem[Aloni et al.(2017)]{Aloni2017}Aloni, D., Blum, K.,  Flauger, R., 2017, JCAP, 05, 017

\bibitem[An et al.(2012)]{An2012} An, T., Wu, F., Yang, J., et al., 2012, ApJS, 198, 5

\bibitem[Angioni et al.(2019)]{Angioni2019} Angioni, R., Ros, E., Kadler, M., et al., 2019, A\&A, 627, A148

\bibitem[Ashoorioon et al.(2021)]{Ashoorioon2019} Ashoorioon, A., Rostami, A., Firouzjaee, J. T., 2021, JHEP, 07, 087

\bibitem[Babak et al.(2016)]{Babak2016}Babak, S., Petiteau, A., Sesana, A., et al., 2015, MNRAS, 455, 1065

\bibitem[Banik et al.(2019)]{Banik2019}Banik, U., van den Bosch, F. C., Tremmel, M., et al., 2019, MNRAS, 483, 1558

\bibitem[Bernal et al.(2017)]{Bernal2017}Bernal, J. L., Bellomo, N., Raccanelli, A., Verde, L., 2017, JCAP, 10, 052

\bibitem[Bird et al.(2016)]{Bird2016}Bird, S., Cholis, I., Mu\~noz, J. B., Ali-Ha\"{\i}moud, Y., Kamionkowski, M., 2016, PRL, 116, 201301

\bibitem[Blandford \& K\"{o}nigl(1979)]{Blandford79} Blandford, R. D., K\"{o}nigl, A., 1979, ApJ, 232, 34

\bibitem[Brandt(2016)]{Brandt2016}Brandt, T. D., 2016, ApJ, 824, L31

\bibitem[Caldwell et al.(1996)]{Caldwell1996}Caldwell, R. R., Chamblin, A., Gibbons, G. W., 1996, PRD, 53, 7103

\bibitem[Cao et al.(2015)]{Cao15}Cao, S., Biesiada, M., Zheng, X., Zhu, Z.-H., 2015, ApJ, 806, 66

\bibitem[Cao et al.(2017a)]{Cao17a}Cao, S., Biesiada, M., Jackson, J., Zheng, X.-G., Zhao, Y.-H., 2017a, JCAP, 02, 012

\bibitem[Cao et al.(2017b)]{Cao17b}Cao, S., Zheng, X.-G., Biesiada, M., et al., 2017b, A\&A, 606, A15

\bibitem[Cao et al.(2018)]{Cao18} Cao, S., Biesiada, M., Qi, J.-Z., et al., 2018, EPJC, 78, 749

\bibitem[Cao et al.(2020)]{Cao20} Cao, S., Qi, J.-Z., Biesiada, M., et al., 2020, ApJL, 888, L25

\bibitem[Carr \& Silk(2018)]{Carr2018}Carr, B., Silk, J., 2018, MNRAS, 478, 3756

\bibitem[Carr et al.(2021)]{Carr2021}Carr, B., Kohri, K., Sendouda, Y., Yokoyama, J., 2021, Rept. Prog. Phys., 84, 116902

\bibitem[Carr \& Hawking(1974)]{Carr1974} Carr, B. J., Hawking, S. W., 1974, MNRAS, 168, 399

\bibitem[Carr(1975)]{Carr1975}Carr, B. J., 1975, ApJ, 201, 1

\bibitem[Carr \& Lidsey(1993)]{Carr1993}Carr, B. J., Lidsey, J. E., 1993, PRD, 48, 543

\bibitem[Carr et al.(1994)]{Carr1994}Carr, B. J., Gilbert, J. H., Lidsey, J. E., 1994, PRD, 50, 4853

\bibitem[Carr et al.(1999)]{Carr1999}Carr, B., J., Sakellariadou, 1999, ApJ, 516, 195

\bibitem[Carr et al.(2016)]{Carr2016}Carr, B. J., Kohri, K., Sendouda, Y., Yokoyama, J., 2016, PRD, 94, 044029

\bibitem[Casadio et al.(2021)]{Casadio2021} Casadio, C., Blinov, D., Readhead, A. C. S., et al., 2021, MNRAS, 507, L6

\bibitem[Chen et al.(2016)]{Chen2016}Chen, L., Huang, Q.-G., Wang, K., 2016, JCAP, 12, 044

\bibitem[Chen \& Cai(2019)]{Cai2019}Chen, C., Cai, Y.-F., 2019, JCAP, 10, 068

\bibitem[Chen et al.(2019)]{Chen2019b}Chen, Z.-C., Yuan, C, Huang, Q.-G., 2020, PRL, 124, 251101

\bibitem[Chen \& Huang(2020)]{Chen2020}Chen, Z.-C., Huang, Q.-G., 2020, JCAP, 08, 039

\bibitem[Clark et al.(2018)]{Clark2018}Clark, S., Dutta, B., Gao, Y., Ma, Y.-Z., Strigari, L. E., 2018, PRD, 98, 043006

\bibitem[Clesse \& Garc\'ia-Bellido(2015)]{Clesse2015}Clesse, S., Garc\'ia-Bellido, J., 2015, PRD, 92, 023524

\bibitem[Clesse \& Garc\'ia-Bellido(2017a)]{Clesse2017a}Clesse, S., Garc\'ia-Bellido, J., 2017a, Phys. Dark Univ. 15, 142

\bibitem[Clesse \& Garc\'ia-Bellido(2017b)]{Clesse2017b}Clesse, S., Garc\'ia-Bellido, J., 2017b, Phys. Dark Univ. 18, 105

\bibitem[De Luca et al.(2020)]{Luca2020}De Luca, V., Franciolini, G., Pani, P., Riotto, A., 2020, JCAP, 06, 044

\bibitem[DeRocco \& Graham(2019)]{DeRocco2019}DeRocco, W., Graham, P. W., 2019, PRL, 123, 251102

\bibitem[Diego(2020)]{Diego2020}Diego, J. M., 2020, PRD, 101, 123512

\bibitem[Duechting(2004)]{Duechting2004}Duechting, N., 2004, PRD, 70, 064015

\bibitem[Fanti et al.(1995)]{Fanti1995} Fanti, C., Fanti, R., Dallacasa, D., Schilizzi, R. T., Spencer, R. E., Stanghellini, C., 1995, A\&A, 302, 317

\bibitem[Fu et al.(2019)]{Fu2019} Fu, C.-J., Wu, P.-X., Yu, H.-W., 2019, PRD, 100, 063532

\bibitem[Graham et al.(2015)]{Graham2015}Graham, P. W., Rajendran, S., Varela, J., 2015, PRD, 92, 063007

\bibitem[Green \& Kavanagh(2021)]{Green2020}Green, A. M., Kavanagh, B. J., 2021, J.Phys.G, 48, 043001

\bibitem[Griest et al.(2013)]{Griest2013}Griest, K., Cieplak, A. M., Lehner, M. J., 2013, PRL, 111, 181302

\bibitem[Gurvits(1994)]{Gurvits94} Gurvits, L. I., 1994, ApJ, 425, 442

\bibitem[Gurvits, Kellerman \& Frey(1999)]{Gurvits99} Gurvits, L. I., Kellerman, K. I., Frey, S., 1999, A\&A, 342, 378

\bibitem[Halder \& Banerjee(2021)]{Halder2020}Halder, A., Banerjee, S., 2021, PhRvD, 103, 063044

\bibitem[Hasegawa \& Kawasaki(2018)]{Hasegawa2018}Hasegawa, F., Kawasaki, M., 2018, PRD, 98, 043514

\bibitem[Hawking(1971)]{Hawking1971}Hawking, S. W., 1971, MNRAS, 152, 75.

\bibitem[Hawking et al.(1982)]{Hawking1982}Hawking, S. W., Moss, I. G., Stewart, J. M., 1982, PRD, 26, 2681

\bibitem[Hawking(1989)]{Hawking1989}Hawking, S. W., 1989, PLB, 231, 237

\bibitem[Hektor et al.(2018)]{Hektor2018}Hektor, A., Hutsi, G., Marzola, L., Raidal, M., Vaskonen, V., Veermae, H., 2018, PRD, 98, 023503

\bibitem[Hogan(1984)]{Hogan1984}Hogan, C. J., 1984, PLB, 143, 87

\bibitem[Huang(2019)]{Huang2019}Huang, Z.-Q., 2019, PRD, 99, 103537

\bibitem[H\"utsi, et al.(2021)]{Gert2020}H\"utsi, G., Raidal, M., Vaskonen, V., Veerm\"ae, H., 2021, JCAP, 03, 068

\bibitem[Jackson(2004)] {Jackson04} Jackson, J. C., 2004, JCAP, 11, 007

\bibitem[Jackson \& Jannetta(2006)] {Jackson06} Jackson, J. C., Jannetta, A. L., 2006, JCAP, 11, 002

\bibitem[Jaffe \& Backer(2003)]{Jaffe2003}Jaffe, A. H., Backer, D. C., 2003, ApJ, 583, 616

\bibitem[Ji et al.(2018)]{Ji2018}Ji, L., Kovetz, E. D., Kamionkowski, M., 2018, PRD, 98, 123523

\bibitem[Kassiola et al.(1991)]{Kassiola1991}Kassiola, A., Kovner, I., Blandford, B. D., 1991, ApJ, 358, 5

\bibitem[Kawasaki et al.(2012)]{Kawasaki2012}Kawasaki, M., Kusenko, A., Yanagida, T. T., 2012, PLB, 711, 1

\bibitem[Kawasaki \& Murai(2019)]{Kawasaki2019}Kawasaki, M., Murai, K., 2019, PRD, 100, 103521

\bibitem[Kellermann(1993)] {Kellermann93} Kellermann, K. I., 1993, Nature, 361, 134

\bibitem[Kitajima \& Takahashi(2020)]{Kitajima2020}Kitajima, N., Takahashi, F., 2020, JCAP, 11, 060,

\bibitem[Kohri et al.(2014)]{Kohri2014}Kohri, K., Nakama, T., Suyama, T., 2014, PRD, 90, 083514

\bibitem[Koushiappas \& Loeb(2017)]{Koushiappas2017}Koushiappas, S. M., Loeb, A., 2017, PRL, 119, 041102

\bibitem[Laha(2019)]{Laha2019}Laha, R., 2019, PRL, 123, 251101

\bibitem[Laha et al.(2020)]{Laha2020a}Laha, R., Mu\~noz, J. B., Slatyer, T. R., 2020, PRD, 101, 123514

\bibitem[Liao et al.(2020)]{Liao2020}Liao, K., Zhang, S.-B., Li, Z., Gao, H., 2020, ApJL, 896, L11

\bibitem[Liao et al.(2020)]{Liao2020a}Liao, K., Tian, S.-X., Ding, X.-H., 2020, MNRAS, 495, 2002

\bibitem[Liu et al.(2021)]{Liu21} Liu, T., Cao, S., Zhang, S., Gong, X., Guo, W., Zheng, C., 2021, EPJC, 81, 903


\bibitem[Malkin(2018)]{Malkin2018} Malkin, Z., 2018, ApJS, 239, 20

\bibitem[Mediavilla et al.(2017)]{Mediavilla2017}Mediavilla, E., Jimenez-Vicente, J.,  Munoz, J. A., VivesArias, H., Calderon-Infante, J., 2017, ApJ, 836, L18

\bibitem[Motohashi et al.(2020)]{Motohashi2020}Motohashi, H., Mukohyama, S., Oliosi, M., 2020, JCAP, 03, 002

\bibitem[Murgia et al.(2019)]{Murgia2019}Murgia, R., Scelfo, G., Viel, M., Raccanelli, A., 2019, PRL, 123, 071102

\bibitem[Nakama et al.(2016)]{Nakama2016} Nakama, T., Suyama, T., Yokoyama, J., 2016, PRD, 94, 103522

\bibitem[Niikura et al.(2019)]{Niikura2019}Niikura, H., Masahiro, T., Naoki, Y., et al., 2019, Nature Astronomy, 3, 524.

\bibitem[Niikura et al.(2019a)]{Niikura12019a}Niikura, H., Takada, M., Yokoyama, S., Sumi, T., Masaki, S., 2019, PRD, 99, 083503

\bibitem[Nemiroff(1989)]{Nemiroff1989}Nemiroff, R. J., 1989, ApJ, 341, 579

\bibitem[Ojha et al.(2005)]{Ojha2005} Ojha, R., Fey, A. L., Charlot, P., et al., 2005, AJ, 130, 2529

\bibitem[Pi et al.(2018)]{Pi2018}Pi, S., Zhang, Y.-L., Huang, Q.-G., Sasaki, M., 2018, JCPA, 1805, 42

\bibitem[Poulin et al.(2017)]{Poulin2017}Poulin, V., Serpico, P. D., Calore, F., Clesse, S., Kohri, K., 2017, PRD, 96, 083524 

\bibitem[Press \& Gunn.(1973)]{Press1973}Press, W. H., Gunn, J. E., 1973, ApJ, 185, 397

\bibitem[Preston et al.(1985)]{Preston85} Preston, R. A., Morabito D. D., Williams J. G., et al., 1985, AJ, 90, 1599

\bibitem[Qi et al.(2019)]{Qi19}Qi, J.-Z., Cao, S., Zhang, S.-X., et al., 2019, MNRAS, 483, 1104  

\bibitem[Qi et al.(2021)]{Qi21} Qi, J.-Z., Zhao, J.-W., Cao, S., Biesiada, M., Liu, Y.-T., 2021, MNRAS, 503, 2179

\bibitem[Readhead et al.(1996)]{Readhead1996} Readhead, A. C. S., Taylor, G. B., Xu, W., Pearson, T. J., Wilkinson, P. N., Polatidis, A. G., 1996, ApJ, 460, 612

\bibitem[Sasaki et al.(2016)]{Sasaki2016}Sasaki, M., Suyama, T., Tanaka, T., Yokoyama, S., 2016, PRL, 117, 061101 

\bibitem[Sasaki et al.(2018)]{Sasaki2018}Sasaki, M., Suyama, T., Tanaka1, T., Yokoyama, S., 2018, GReGr, 35, 063001

\bibitem[Serpico et al.(2020)]{Serpico2020}Serpico, P. D., Poulin, V., Inman, D., Kohri, K., 2020, Phys.Rev.Res., 2, 023204

\bibitem[Sesana et al.(2008)]{Sesana2008}Sesana, A., Vecchio, A., Colacino, C. N., 2008, MNRAS, 390, 192

\bibitem[Sesana et al.(2009)]{Sesana2009}Sesana, A., Vecchio, A., Volonteri, M., 2009, MNRAS, 394, 2255

\bibitem[Shinohara et al.(2021)]{Shinohara2020}Shinohara, T., Suyama, T., Takahashi, T., 2021, PRD,  104, 023526

\bibitem[Sokolovsky et al.(2011)]{Sokolovsky2011} Sokolovsky, K. V., Kovalev, Y. Y., Pushkarev, A. B., Mimica, P., Perucho, M., 2011, A\&A, 535, A24

\bibitem[Tisserand et al.(2007)]{Tisserand2007}Tisserand, P., Guillou, P.,  Afonso, C., et al., 2007, A\&A 469, 387

\bibitem[Turner et al.(1984)]{Turner1984}Turner, E. L., Ostriker, J. P., Gott, J. R., 1984, ApJ, 284, 1

\bibitem[Urrutia \& Vaskonen (2021)]{Urrutia2021}Urrutia, J., Vaskonen, V., 2021, MNRAS, 509, 1358

\bibitem[Wang et al.(2018)]{Wang2018}Wang, S., Wang, Y.-F.,  Huang, Q.-G., Li,T. G. F., 2018, PRL, 120, 191102

\bibitem[Wilkinson et al.(2001)]{Wilkinson2001}Wilkinson, P. N., Henstock, D. R., Browne, W. A., et al., 2001, PRL, 86, 4

\bibitem[Woods et al.(2019)]{Woods2019}Woods, T. E., Agarwal, B., Bromm, V., et al., 2019, PASA, 36, e027

\bibitem[Xu et al.(2018)]{Xu18} Xu, T. P., Cao, S., Qi, J.-Z., Biesiada, M., Zheng, X., Zhu, Z.-H., 2018, JCAP, 06, 042

\bibitem[Yang et al.(2020)]{Yang2020} Yang, J.,  Wang, F., Fan, X., et al., 2020, ApJ, 897, L14

\bibitem[Zheng et al.(2017)]{Zheng17} Zheng, X., Biesiada, M., Cao, S., Qi, J.-Z., Zhu, Z.-H., 2017, JCAP, 10, 030

\bibitem[Zheng et al.(2020)]{Zheng20} Zheng, X., Liao, K., Biesiada, M., Cao, S., Liu, T., Zhu, Z.-H., 2020, ApJ, 892, 103

\bibitem[Zhou et al.(2021)]{Zhou2021}Zhou, H., Li, Z.-X., Liao, K., Niu, C.-H., Gao, H., Huang, Z.-Q., Huang, L., Zhang, B., 2021, arXiv: 2109.09251

\bibitem[Zhou et al.(2022)]{Zhou2022} Zhou, H., Li, Z.-X., Huang, Z.-Q., Gao, H., Huang, L., 2022, MNRAS, 511, 1141

\bibitem[Zhu et al.(2014)]{Zhu2014}Zhu, X.-J., Hobbs, G., Wen, L. et al., 2014, MNRAS, 444, 3709

\bibitem[Zumalacarregui \& Seljak(2018)]{Zumalacarregui2018} Zumalacarregui, M., Seljak, U., 2018, PRL, 121, 141101


\end{thebibliography}
\end{document}